\documentclass[10pt]{article}
\usepackage{CJK}
\usepackage{amsmath,amssymb}
\usepackage{graphicx,subfigure}
\usepackage{bm,cite,url}
\usepackage{color}
\usepackage[labelfont=bf,labelsep=period,justification=raggedright]{caption}

\begin{document}

\title{Quantitative Function and Algorithm for Community Detection in Bipartite Networks}

\author{Zhenping Li$^{1}$, Rui-Sheng Wang$^2$, Shihua Zhang$^{3\ast}$, Xiang-Sun Zhang$^{3\ast}$}

\date{\small 1. School of Information, Beijing Wuzi University, Beijing, China;\\
2. Department of Medicine, Brigham and Women's Hospital, Harvard Medical School, Boston, USA;\\
3. National Center for Mathematics and Interdisciplinary Sciences, Academy of Mathematics and Systems Science, CAS, Beijing, China;\\
$\ast$ Corresponding author. E-mail: zsh@amss.ac.cn, zxs@amt.ac.cn}

\maketitle

\begin{abstract}
Community detection in complex networks is a topic of high interest in many fields. Bipartite networks are a special type of complex networks in which nodes are decomposed into two disjoint sets, and only nodes between the two sets can be connected. Bipartite networks represent diverse interaction patterns in many real-world systems, such as predator-prey networks, plant-pollinator networks, and drug-target networks. While community detection in unipartite networks has been extensively studied in the past decade, identification of modules or communities in bipartite networks is still in its early stage. Several quantitative functions proposed for evaluating the quality of bipartite network divisions are based on null models and have distinct resolution limits. In this paper, we propose a new quantitative function for community detection in bipartite networks, and demonstrate that this quantitative function is superior to the widely used Barber's bipartite modularity and other functions. Based on the new quantitative function, the bipartite network community detection problem is formulated into an integer programming model. Bipartite networks can be partitioned into reasonable overlapping communities by maximizing the quantitative function. We further develop a heuristic and adapted label propagation algorithm (BiLPA) to optimize the quantitative function in large-scale bipartite networks. BiLPA does not require any prior knowledge about the number of communities in the networks. We apply BiLPA to both artificial networks and real-world networks and demonstrate that this method can successfully identify the community structures of bipartite networks.

\end{abstract}

\section{Introduction}
Many real-world systems, such as the Internet, social relationships, food webs, and biological systems, can be represented as complex networks \cite{Albert:02,Newman:03, Hu2008, Zhang2007}. Community and its detection is an important and valuable topic of great interest in network science and many other disciplines from sociology, computer science to biology \cite{Fortunato2010,Newman2012,Zhang2007,Zhang2012}. A community in a complex network is often defined as a set of nodes which are densely connected with each other, but sparsely connected with nodes outside of the set.

To evaluate the quality of a network partition or community structure, a proper quantitative function is needed. Modularity $Q$ introduced by Newman and Girvan \cite{Newman2004} is the most widely used function. $Q$ measures the density of edges within communities as compared to a null model. In other words, a partition with high modularity should be the one that the density of edges in modules is significantly higher than random expectation. Modularity optimization has become a very popular method for community detection in the past decade. However, $Q$ has been exposed to have serious resolution limit issues \cite{Fortunato2007,Rosvall2007}. It contains an intrinsic scale that depends on the total number of edges in the network which makes it fail to detect dense communities smaller than this scale \cite{Fortunato2007}. Recently, Bagrow \cite{Bagrow2012} reported that trees and treelike networks can have arbitrarily high values of modularity $Q$ which contradicts the notion of communities as being unusually densely interconnected groups of nodes.

To overcome the above issues, several other quantitative functions have been proposed \cite{Li2008, Zhang2010, Ahn2010, Li2013}. For example, Li et al. \cite{Li2008} proposed community density $D$ based on the concept of average degree of a community. Ahn et al. \cite{Ahn2010} proposed the link community detection concept and defined a quantitative function based on the average link density. Li et al. \cite{Li2013} improved Ahn's link community quantitative function. Although these quantitative functions \cite{Li2008, Zhang2010, Ahn2010, Li2013} show certain advantages compared to the modularity $Q$, they are only designed for detecting communities in unipartite networks.

Although most attention has focused on the community detection in unipartite networks (see \cite{Fortunato2010} and references therein), many real-world relations, such as plant-pollinator, order-item, paper-author, and event-attendee, are more suitable to be represented as bipartite networks \cite{Guimer2007}. In bipartite networks, nodes are divided into two disjoint sets (i.e., a bipartite network is composed of two types of nodes), and only two nodes from different sets can be connected. In the past several years, community detection in bipartite networks has attracted great interests as well \cite{Guimer2007, Zan2011, Costa2011, Costa2014, Liu2009, Liu2010, Lee2013, Mao2012, Murata2010}.

Several quantitative functions and algorithms for community detection in bipartite networks have been developed \cite{Freeman2003, Guimer2007, Barber2007, Murata2010}. Freeman \cite{Freeman2003} proposed a projection-based algorithm for bipartite community detection, where a bipartite network is projected to a unipartite network. In the projected network, two nodes are connected if they are adjacent to one or more common nodes in the bipartite network. Obviously, some information of the original bipartite network is lost in the projected network, even if the projection is weighted.

In Guimer\`{a}'s opinion, a bipartite network community is composed of nodes of the same type \cite{Guimer2007}. In order to measure the connectivity between nodes of different types, a bipartite modularity function considering both node types has to be used. Barber proposed a bipartite modularity function \cite{Barber2007, Barber08} based on the assumption that a community is a bipartite subgraph composed of nodes of both types. Murata and Ikeya \cite{Murata2010} proposed another bipartite modularity that uses a pair of two values to represent bipartite modularity in both directions, which allows one-to-many correspondence of communities of different node types. This bipartite modularity can reveal results consistent with those obtained by Newman's modularity when it is applied to unipartite networks. Suzuki and Wakita \cite{Suzuki2009} modified the version of Murata's bipartite modularity so as to reflect the multi-facet correspondence among communities. In addition, several studies have designed algorithms to maximize these modularity functions for community detection in bipartite networks \cite{Liu2009, Murata2009, Costa2014, Murata2010, Liu2010}. More recently, Chang and Tang proposed a probabilistic model to find modules in unipartite, bipartite, and mixture networks \cite{Chang2014}.

So far there has been no agreement on a standard definition for community detection in bipartite networks. Since the above definitions of bipartite modularity are all based on the null model used in Newman-Girvan modularity, they all have the resolution limit issue \cite{Murata2010,Fortunato2007} and fail to detect communities smaller than a detectable scale that depends on the size of a network and the interconnectivity of its communities. When the scale of the network is sufficiently large, optimizing bipartite modularity favors network divisions with groups of small communities merged into larger communities, which may lead to ambiguities \cite{Liu2009}.

In this paper, we propose a novel quantitative function -- bipartite partition density for evaluating community partitions in (both unweighted and weighted) bipartite networks. This quantitative function is not a simple generalization of the average link community density in \cite{Li2013} for bipartite networks. In this new quantitative function, we partition the nodes rather than links of bipartite networks into communities, i.e., every node must belong to at least one community while some links may not belong to any community. We formulate the community partition problem for a bipartite network into an integer programming model. In addition to the simple form, we show that the proposed criterion overcomes the resolution limit for community detection in bipartite networks based on theoretical analysis and numerical tests on artificial and real-world networks. Then we design a heuristic algorithm (BiLPA) for efficient community detection in large-scale bipartite networks. We demonstrate their effectiveness with applications onto artificial and real-world bipartite networks.


\section{Partition Density of Bipartite Networks}
A bipartite network $G=(U,V,E)$ is a graph with two disjoint types of nodes $U$ and $V$, $U \cap V =\emptyset$, such that there is no edge connecting two nodes of the same type. Let $L(U,V)$ be the number of edges with one node in $U$ and the other node in $V$, then $L(U,V)=|E|$ is the total number of edges in bipartite network $G$.

A complete bipartite network or a biclique $G=(U,V,E)$ is a bipartite network, with each node in $U$ adjacent to each node in $V$. In other words, there are $L(U,V)=|U|\times |V|$ edges in the complete bipartite network. Let $B(m,n)$ denote a biclique with $m$ nodes in $U$ and $n$ nodes in $V$. $B(m,n)$ has the maximal number of edges among all bipartite networks with the same node sets $U$ and $V$.

{\bf Definition 1. Bipartite graph density} The density $D(G)$ of a bipartite network $G=(U,V,E)$ is defined as the ratio of the number of edges in $G$ to the number of edges of a corresponding biclique $B(|U|,|V|)$, which can be calculated by $D(G)=\frac{L(U,V)}{|U|\times |V|}$. Obviously, for any bipartite network without parallel edges, the maximal graph density value is 1 when it is a biclique, and the minimal graph density value is 0 when it is an empty bipartite network.

{\bf Definition 2. Community partition density} Given a bipartite network $G=(U,V,E)$, we assume that $G=\{G_1,\cdots,G_K\}$ is a partition of $G$ into $K$ subgraphs, $G_c=(U_c,V_c,E_c)$, $c=1,2,\cdots,K$, then $\cup_{c=1}^K U_c=U$, $\cup_{c=1}^K V_c=V$, and $E_c=\{e_{ij}\in E|u_i \in U_c, v_j \in V_c\}$. Let $L(U_c,V_c)=|E_c|$ be the number of edges with one node in $U_c$ and the other node in $V_c$, then the density of subgraph $G_c$ is defined as
$$D_c=\frac{L(U_c,V_c)}{|U_c|\times |V_c|}.$$
The partition density $D$ of a bipartite network is defined as the weighted average of $D_c$, where the weight coefficient is the ratio of the number of edges in each community relative to the total number of edges in bipartite network $G$:

\begin{equation}
D=\sum_{c}{\frac{L(U_c,V_c)}{L(U,V)}\cdot D_c}=\frac{1}{L(U,V)}\sum_{c}{\frac{L(U_c,V_c)^2}{|U_c|\times |V_c|}}.\label{eq(1)}
\end{equation}

We can see that the maximum value of partition density $D$ is 1 when each community is a biclique and there is no edge between the communities, and the minimum value of $D$ is 0 when each community is an empty bipartite graph. If a bipartite network is partitioned into overlapping communities, we can also use formula \ref{eq(1)}  to calculate the partition density.

We can easily extend this definition for weighted bipartite networks. Let $G=(U,V,E,W)$ be a weighted bipartite network, where $W=(w_{ij})_{p\times q}$ is the weight matrix, and $w_{ij}$ is the weight of edge $e_{ij}$ taking values in [0,1] without loss of generality. Let $W(U,V)$ be the sum of all edges' weight between node sets $U$ and $V$, then we can define the weighted partition density $D_W$ of weighted bipartite network $G=(U,V, E, W)$ as follows:
$$D_W=\sum_{c}{\frac{W(U_c,V_c)}{W(U,V)}\cdot D_c}=\frac{1}{W(U,V)}\sum_{c}{\frac{W(U_c,V_c)^2}{|U_c|\times |V_c|}}.\quad \quad $$
By maximizing the partition density $D$ and $D_W$, we can find the optimal community partition of unweighted and weighted bipartite networks respectively.

{\bf Definition 3. Core degree} Given a partition of bipartite network $G=\{G_1,\cdots,G_K\}$, for any node $u_i \in U$ and any community $G_c=(U_c,V_c,E_c)$, the core degree of node $u_i$ to community $G_c$ is defined as $CD(u_i,G_c)=\frac{|N(u_i)\cap V_c|}{|V_c|}$; similarly, for any node $v_j \in V$ and the community $G_c$, the core degree of node $v_j$  to community $G_c$ is defined as $CD(v_j,G_c)=\frac{|N(v_j)\cap U_c|}{|U_c|}$, where $N(u_i)$ and $N(v_j)$ denote the set of nodes adjacent to $u_i$ or $v_j$ in bipartite network $G$.

If a node has a high core degree to its community, the node might be an important member of the community, so we call it a core node. If a node has a low core degree to its community, it might not be an important member of the community, thus we call it a peripheral node. In a biclique, every node has core degree 1, and is a core node. A sparse bipartite network has plenty of peripheral nodes.

Given a partition $G=\{G_1,\cdots,G_K\}$ of bipartite network $G$, for a node $u_i \in U$ which does not belong to any community. If we put node $u_i$ to community $G_c=(U_c,V_c,E_c)$, then the partition density of the network will increase by
\begin{equation}
\begin{array}{l}
\displaystyle \vartriangle D(u_i, G_c)=\frac{1}{L(U,V)}[\frac{(L(U_c,V_c)+|N(u_i)\cap V_c|)^2}{(|U_c|+1)|V_c|}-\frac{L(U_c,V_c)^2}{|U_c||V_c|}]\\
\displaystyle  \\
\displaystyle \approx\frac{1}{L(U,V)}[\frac{(L(U_c,V_c)+|N(u_i)\cap V_c|)^2}{|U_c||V_c|}-\frac{L(U_c,V_c)^2}{|U_c||V_c|}]\\
\displaystyle  \\
\displaystyle = \frac{1}{L(U,V)}\frac{(2L(U_c,V_c)+|N(u_i)\cap V_c|)|N(u_i)\cap V_c|}{|U_c||V_c|}\\
\displaystyle  \\
\displaystyle = \frac{1}{L(U,V)}\frac{(2L(U_c,V_c)+|N(u_i)\cap V_c|)}{|U_c|}\frac{|N(u_i)\cap V_c|}{|V_c|}\\
\displaystyle  \\
\displaystyle = \frac{1}{L(U,V)}\frac{(2L(U_c,V_c)+|N(u_i)\cap V_c|)}{|U_c|}CD(u_i,G_c)\\
\end{array} \label{eq2}
\end{equation}

\begin{equation}
\begin{array}{l}
\displaystyle \vartriangle D(v_j,G_c)=\frac{1}{L(U,V)}[\frac{(L(U_c,V_c)+|N(v_j)\cap U_c|)^2}{|U_c|(|V_c|+1)}-\frac{L(U_c,V_c)^2}{|U_c||V_c|}]\\
\displaystyle  \\
\displaystyle \approx\frac{1}{L(U,V)}[\frac{(L(U_c,V_c)+|N(v_j)\cap U_c|)^2}{|U_c||V_c|}-\frac{L(U_c,V_c)^2}{|U_c||V_c|}]\\
\displaystyle  \\
\displaystyle = \frac{1}{L(U,V)}\frac{(2L(U_c,V_c)+|N(v_j)\cap U_c|)|N(v_j)\cap U_c|}{|U_c||V_c|}\\
\displaystyle  \\
\displaystyle = \frac{1}{L(U,V)}\frac{(2L(U_c,V_c)+|N(v_j)\cap U_c|)}{|V_c|}\frac{|N(v_j)\cap U_c|}{|U_c|}\\
\displaystyle  \\
\displaystyle = \frac{1}{L(U,V)}\frac{(2L(U_c,V_c)+|N(v_j)\cap U_c|)}{|V_c|}CD(v_i,G_c)\\
\end{array} \label{eq3}
\end{equation}

From the equation (\ref{eq2}) (or (\ref{eq3})) described above, we can see that, if we put a node $u_i$ (or $v_j$) into a community $G_c$ with high core degree, then the network's partition density will increase much. When there are two communities to which node $u_i$ (or $v_j$) has the same core degree, we should put it into the one that contains its most adjacent nodes $|N(u_i) \cap V_c|$ (or $|N(v_j) \cap U_c|$) in order to obtain larger partition density. According to this relationship, we will design a heuristic algorithm (BiLPA) for community detection in bipartite networks.

\section{Community Partition Density Improves Resolution Limits}

Although Barber's bipartite modularity has been widely used for community detection in bipartite networks, it may fail to detect communities smaller than a scale even in case where communities are bicliques. The proposed partition density $D$ can overcome such an issue to some extent. In the following, we will give some examples to demonstrate this point.

\begin{figure}[t]
\begin{center}
\includegraphics[width=0.99\textwidth]{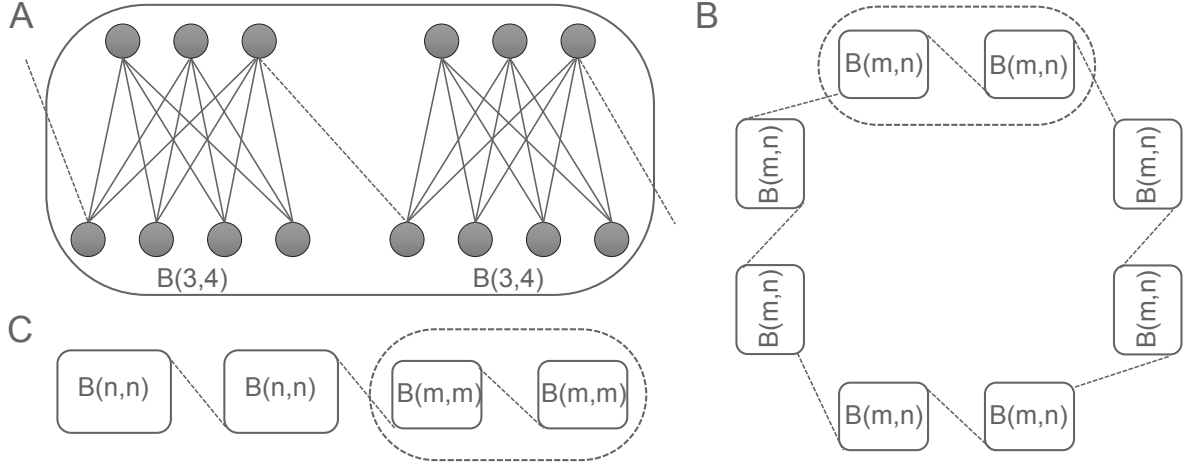}
\caption{(A) The connection relationship of two bicliques in (B). (B) The ring of bicliques. Each square indicates a biclique of size $m+n$, and two adjacent bicliques are connected by one edge. (C) A bipartite network consists of four bicliques, two of which are $B(n,n)$ and the other two are $B(m,m)$. } \label{figure1}
\end{center}
\end{figure}

{\bf A ring of bicliques} A ring of bicliques is a bipartite graph composed of $s$ bicliques connected through single edges (see Figure \ref{figure1}A and \ref{figure1}B). Each biclique is a complete bipartite graph $B(m,n)$ with $m$ nodes in $U$, $n$ nodes in $V$ ($m,n \geq 2$) and $mn$ edges. The network has a total of $(m+n)s$ nodes and $(mn+1)s$ edges.

The network has a clear community structure where each biclique corresponds to a single community, but the correct result cannot be obtained by optimizing Barber's bipartite modularity $Q$. If we partition the network into $s$ communities, then each community corresponds to a single biclique. If we partition the network into $\frac{s}{2}$ communities, then each community consists of two adjacent bicliques. When $s>2mn+1$, we can easily see that Barber's bipartite modularity for these two partitions: $Q_{s}<Q_{\frac{s}{2}}$ (see Appendix A). Thus, by maximizing Barber's bipartite modularity, we cannot partition the network into reasonable communities.

Assuming that $s\geq 4$, and it can be exactly divided by an integer $k$, $k\geq 2$, now we calculate the partition density $D_{s}$, $D_{\frac{s}{2}}$, $D_{\frac{s}{3}}$ by dividing the network into $s$, $\frac{s}{2}$ and $\frac{s}{3}$ communities, respectively. We find that $D_{s}>D_{\frac{s}{2}}$ and $D_{s}>D_{\frac{s}{3}}$ with $mn\geq 2$, $s\geq 4$, which confirm that the partition density $D_{s}$ has larger value.

Although the above analysis is conducted for the special partition that the $k$ consecutive bicliques are considered as a single community, by similar argument we can deduce that such a result is actually valid for any kind of grouping bicliques (i.e. any combination of bicliques as communities). In fact, the partition density of any other combination of bicliques as communities is less than that of the special partition $D_{s}$ mentioned above. Therefore, these results, along with the fact that the maximum value of $D$ is obtained in complete bipartite networks lead to a conclusion that the maximum value of $D$ corresponds to the correct partition with each single biclique as a community.

{\bf A bipartite network consisting of bicliques of different sizes} Assume that there is a bipartite network consisting of four bicliques, two of which are $B(n,n)$, and the other two are $B(m,m)$, for $2\leq m\leq n $ (see Figure 1C). This bipartite network has a natural community structure of four communities, where each biclique corresponds to one community. If we partition the bipartite network by optimizing Barber's bipartite modularity $Q$, we will obtain three communities, with two small bicliques merged into one community (Figure 1C).

Let $Q_{separate}$ and $D_{separate}$ be Barber's bipartite modularity and our partition density when the network is partitioned into four communities corresponding to the four bicliques in Figure 1C, and $Q_{merge}$ and $D_{merge}$ denote Barber's bipartite modularity and our partition density when two small bicliques merge into one community. We can see that $Q_{separate}<Q_{merge}$ with $n\geq m^2+1$, and the two small communities will merge into one community  (Appendix A). In contrast, it is easy to see that $D_{separate}>D_{merge}$ with $m\geq 2$, which indicates the partition density $D$ based optimization does not have the resolution problem mentioned above (Appendix A).

The above analysis is conducted for the special partition wherein two smaller bicliques are merged into a community with each other biclique as a separate community. With the fact that bicliques have maximum partition density, it is easy to see that any other partition has a lower partition density than the one with each biclique as a community. Therefore, the optimal value of partition density $D$ corresponds to the correct partition. In contrast to bipartite modularity $Q$, optimizing partition density $D$ can correctly detect communities of any size. In summary, optimization of partition density $D$ can often achieve correct partitions while Barber's bipartite modularity not. Thus, we deduce that bipartite partition density $D$ can act as a proper quantitative function for community detection in bipartite networks.

\section{Integer Programming Model for Bipartite Community Partition}
Let $U=\{u_1,u_2, \cdots,u_p\}$, $V=\{v_1,v_2,\cdots,v_q\}$ be two disjoint node sets of bipartite network $G=(U,V,E)$, and $A=(a_{i \times j})_{p \times q}$ be the adjacency matrix of node sets $U$ and $V$ in $G$, where $a_{ij}=1$ indicates there is an edge $e_{ij}$ between nodes $u_i$ and $v_j$, $a_{ij}=0$ otherwise. Note that $|E|=\sum_{i=1}^p \sum_{j=1}^q a_{ij}$ is the total number of edges in $G$. Let's define binary variables $x_{ic}$, $y_{jc}$ to represent whether node $u_i$ and node $v_j$ belong to community $G_c$ or not, respectively:
\begin{equation*}
 x_{ic}=\left\{ \begin{array}{ll}1& \quad  \mbox{if}~ u_i \in G_c\\
 0& \quad \mbox{otherwise,}
\end{array}
\right.
\end{equation*}

\begin{equation*}
y_{jc}=\left\{ \begin{array}{ll}1& \quad  \mbox{if}~ v_j\in G_c\\
0& \quad \mbox{otherwise}.
\end{array}
\right.
\end{equation*}

We can formulate the problem of partitioning bipartite network $G$ into $K$ communities into an integer programming model (Model-1):
\begin{equation*}
\begin{array}{ll}\max
\displaystyle D=\frac{1}{|E|}\sum_{c=1}^{K}\frac{(\sum_{i=1}^p \sum_{j=1}^q a_{ij}x_{ic}y_{jc} )^2}{(\sum_{i=1}^px_{ic} )(\sum_{j=1}^qy_{jc}) } \quad \quad \quad \quad \quad\quad \quad\quad\quad\quad\quad\quad\quad\quad(4)
\end{array}
\end{equation*}

\begin{equation*}
\quad s.t.\left\{ \begin{array}{ll}
\displaystyle \sum_{c=1}^{K} x_{ic}= 1 \quad i=1,2,\cdots,p & \quad\quad\quad\quad\quad\quad\quad\quad(5) \\
\displaystyle \sum_{c=1}^{K} y_{jc}= 1 \quad j=1,2,\cdots,q & \quad\quad\quad\quad\quad\quad\quad\quad(6) \\
\displaystyle x_{ic}\in\{ 0,1\}; \quad  i=1,2,\cdots,p, c=1, 2, \cdots, K & \quad\quad\quad\quad\quad\quad\quad\quad(7)\\
\displaystyle y_{jc}\in\{ 0,1\}; \quad j=1,2,\cdots,q, c=1, 2, \cdots, K &  \quad\quad\quad\quad\quad\quad\quad\quad(8)\\
\end{array}
\right.
\end{equation*}
where the objective function (4) is to maximize partition density $D$, constraints (5) and (6) indicate that each node must belong to one and only one community, and constraints (7) and (8) ensure that the variables are binary ones. We can easily solve the integer programming model by using the Lingo software \cite{Scharge2004} for small-scale bipartite networks to see if the model can find communities properly.
To partition a bipartite network into overlapping communities, we can simply relax the constraints (5) and (6) to allow a node to belong to more than one community by changing the constraints to $\sum_{c=1}^{K} x_{ic}\geq 1$ for $i=1,2,\cdots,p$ and $\sum_{c=1}^{K} y_{jc}\geq 1$ for $j=1,2,\cdots,q$. We represent such a model as Model-2.

\begin{figure}[t]
\begin{center}
\includegraphics[width=0.99\textwidth]{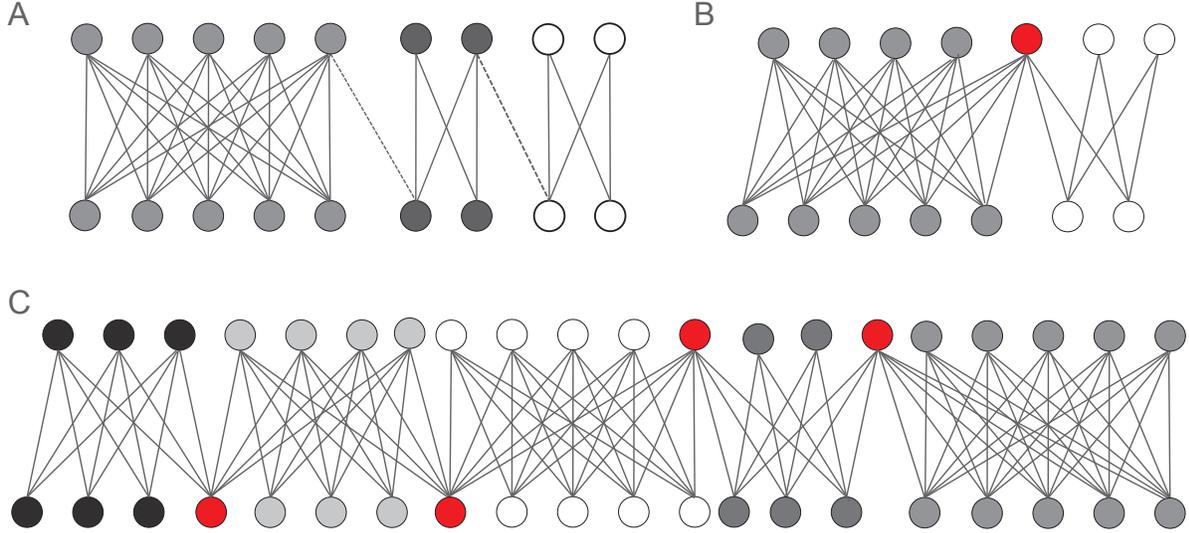}
\caption{(A) Three communities of a bipartite network. The nodes in the same community are drawn in the same color, and the edges between communities are dotted. (B) Partitioning a bipartite network into two overlapping communities. The red node is overlapped by the two communities. (C) A chain of heterogeneous bicliques consisting of bicliques B(3,4), B(4,5), B(5,5), B(4,3) and B(6,5). Two adjacent bicliques are overlapping with one red node. } \label{figure2}
\end{center}
\end{figure}

For example, the bipartite network in Figure \ref{figure2}A has graph density 0.4321. We can easily partition it into two and three communities by solving Model-1 and obtain the partition density 0.6391 and 0.9429. If we partition it into more than 3 communities, the partition density will decrease and thus the maximum partition density is 0.9429 which seems reasonable. For the bipartite network in Figure \ref{figure2}B, we can partition it into two overlapping communities by solving Model-2 with the maximum partition density 1.

\section{BiLPA for Bipartite Community Partition}
To efficiently solve the community partition problem, we further develop a heuristic algorithm (BiLPA) for community detection in large-scale bipartite networks, which does not need to pre-specify the number of communities $K$. Raghavan et al. \cite{Raghavan2007} firstly proposed a label propagation algorithm (LPA) for community detection in large-scale unipartite networks. Barber and Clark \cite{Barber2009} generalized it (LPAb) to optimize bipartite modularity. Liu and Murata \cite{Liu2009,Liu2010} further introduced an improved label propagation algorithm (LPAb+) for this task. However, both LPAb and LPAb+ are designed for maximizing Barber's bipartite modularity. In this paper, we introduce a revised Label Propagation Algorithm (BiLPA) for community detection in large-scale bipartite networks by maximizing our partition density $D$.

Let $A=(a_{ij})_{p \times q}$ be the adjacency matrix of bipartite network $G$, and $d_{u_i}$ and $d_{v_j}$ be the degrees of node $u_i$ and node $v_j$, respectively. The key idea of BiLPA is as follows. At first, each node $u_i\in U$ in the bipartite network is initially assigned a unique label, indicating its community membership (without causing confusion, we use $r_i$ to indicate node $u_i$'s label.). Then each node $v_j\in V$ acquires a label from its neighboring nodes in $U$, indicating its community membership (we use $b_j$ to indicate the node $v_j$'s label). In the following steps, the labels of nodes in $U$ and nodes in $V$ are updated in turn, until a local maximum partition density is obtained. The updating rule is based on the following idea: In order to find the maximum partition density, every node always chooses to join in a community that leads to the maximum increase of the partition density. From equations (\ref{eq2}) and (\ref{eq3}), we can see that the increase of the partition density after node $u_i$ (or $v_j$) joins in community $G_c=(U_c,V_c,E_c)$ is related to the core degree $CD(u_i,G_c)$ (or $CD(v_j, G_c)$) and the number of node $u_i$'s (or $v_j$'s) neighbors belonging to community $G_c$. So we can give two updating rules as follows: (I) If there is only one community to which the node has the maximum core degree, then the node joins in this community; (II) If there are two or more communities to which the node has maximum core degree, then the node chooses to join the one that contains most nodes adjacent to it.
These rules can be described by the following formulae respectively.
For node $v_j$,

\begin{equation*}
\displaystyle
LB(j)=\underset{g}{\operatorname{arg\,max}}\frac{\sum_{i=1}^pa_{ij}\delta(g,r_i)}{\sum_{i=1}^p\delta(g,r_i)}. \quad \quad \quad\quad\quad\quad\quad\quad(9)
\end{equation*}

\begin{equation*}
\displaystyle
b_j^{new}=\underset{g \in LB(j)}{\operatorname{arg\,max}}{\sum_{i=1}^pa_{ij}\delta(g,r_i)}. \quad \quad  \quad\quad\quad\quad\quad\quad\quad\quad(10)
\end{equation*}

For node $u_i$,

\begin{equation*}
\displaystyle LR(i)=\underset{g}{\operatorname{arg\,max}}\frac{\sum_{j=1}^qa_{ij}\delta(g,b_j)}{\sum_{j=1}^q\delta(g,b_j)}. \quad \quad \quad\quad\quad\quad\quad\quad(11)
\end{equation*}

\begin{equation*}
\displaystyle r_i^{new}=\underset{g \in LR(i)}{\operatorname{arg\,max}} {\sum_{j=1}^qa_{ij}\delta(g,b_j)}, \quad \quad \quad\quad\quad\quad\quad\quad\quad\quad(12)
\end{equation*}
where $g\in\{1,2,..., K\}$ is a community label, and $\delta(x,y)=1$ if $x=y$, otherwise $\delta(x,y)=0$. $LB(j)$ and $LR(i)$ are the label sets of communities corresponding to the maximum core degree. Formula (9) (or (11) finds the label set of communities to which $v_j$ (or $u_i$) has maximum core degree. Formula (10) (or (12)) finds the label of the community among $LB(j)$ (or $LR(i)$) where $v_j$ (or $u_i$) has most adjacent nodes in it. Formulae (9) (or (11)) and (10) (or (12)) mean that a node should be assigned to a community to which the node has maximum core degree and most adjacent nodes.

The BiLPA pseudo-code is presented in {\bf Algorithm BiLPA}. \\

\noindent {\bf Algorithm BiLPA:}
\begin{itemize}
\item Input\\
Input the adjacency matrix $A$ of bipartite network $G=(U,V,E)$, and calculate the number of nodes $p$ ($p=|U|$) in $U$  and the number of nodes $q$ ($q=|V|$) in $V$, respectively. Set the maximum number of iterations $Iter_{\max}$ and the parameter value $\theta$.

\item Output\\
   Output a community partition and its partition density, and the rearranged matrix according to the partition.

\item Step 1. Initialize the labels of nodes in $U$. For each node $u_i$, let $r_i=i$ , $t=0$ and  $D(0)=0$;

\item Step 2. Calculate the number of nodes in $U$ with the same labels. For each label $k$, calculate the number of nodes in $U$ labeled with $k$ using $R(k)=\sum_{i=1}^p \delta(k,r_i)$.

\item Step 3. Update the labels of nodes in $V$. For each node $v_j$, update its label with

$$LB(j)=
    \underset{k}{\operatorname{arg\,max}}\frac{\sum_{i=1}^pa_{ij}\delta(k,r_i)}{R(k)}.$$

$$b_j^{new}=
    \underset{k \in LB(j) }{\operatorname{arg\,max}}{\sum_{i=1}^pa_{ij}\delta(k,r_i)}.$$

\item Step 4. Calculate the number of nodes in $V$ with the same labels. For each label $k$, calculate the number of nodes in $V$ labeled with $k$ using $B(k)=\sum_{j=1}^q \delta(k,b_j)$.

\item Step 5. Update the labels of nodes in $U$. For each node $u_i$, update its label with $$LR(i)=\underset{k}{\operatorname{arg\,max}}\frac{\sum_{j=1}^qa_{ij}\delta(k,b_j)}{B(k)}.$$

$$r_i^{new}=\underset{k \in LR(i)}{\operatorname{arg\,max}} {\sum_{j=1}^qa_{ij}\delta(k,b_j)}.$$

\item Step 6. Calculate the partition density D(t). If $D(t) \leq D(t-1)$ or $t<Iter_{\max}$, go to Step 7, else $t=t+1$, go to Step 2.

\item Step 7. Calculate the final label of each node and the corresponding partition density. For each node $u_i$ in $U$ (or similarly for node $v_j$ in $V$), calculate its core degree $CD(u_i,k)$ to each community $k$ and find its maximum core degree $\max_g CD(u_i,g)$. Calculate the core degree rate $RCD(u_i,k)$ with $RCD(u_i,k)=\frac{CD(u_i,k)}{\max_g CD(u_i,g)}$. For each community label $k$, determine the community membership of $u_i$ by examining whether $RCD(u_i,k)$ is larger than $\theta$ or not, and assign the label to node $u_i$ in $U$ if so. Then calculate the final partition density.

\item Step 8. Rearrange the adjacency matrix $A$ according to the partition result. First, construct a matrix $B$ by adjusting the order of the rows of matrix $A$ according to the label of nodes in $U$. For $k=1,\cdots,p$, let the $R(k)$ rows of $A$ corresponding to nodes in $U$ labeled with $k$ consist of the $R(1)+\cdots +R(k-1)+1, R(1)+\cdots +R(k-1)+2,\cdots, R(1)+\cdots +R(k-1)+R(k)$ rows of matrix $B$, respectively. Then, construct the rearranged matrix $C$ from $B$ according to the label of nodes in $V$. For $k=1,\cdots ,p$, let the $R(k)$ columns of $B$ corresponding to nodes in $V$ labeled with $k$ (matrices $B$ and $A$ have the same column labels) consist of the $R(1)+\cdots +R(k-1)+1, R(1)+\cdots +R(k-1)+2,\cdots, R(1)+\cdots +R(k-1)+R(k)$ columns of matrix $C$, respectively.

\end{itemize}

We allow a node to belong to multiple communities in Step 7 which enables us to find overlapping nodes in a bipartite network.

In BiLPA, initialization of each node in $U$ with unique labels requires $O(p)$ time. At each iteration, for each node $u_i$ (or $v_j$), we first group its neighbors according to their labels ($O(q)$ or $O(p)$), then calculate the core degree to each community ($O(q)$ or $O(p)$), and select the community corresponding to its maximum core degree and containing its largest number of neighbors ($O(q)$ or $O(p)$). Finally, we calculate the partition density ($O(pq)$), where $p$ and $q$ are the numbers of two disjoint sets of nodes in the bipartite network. So the total time complexity of each iteration is $O(pq)$.

\section{Experiment Results}

\subsection{A chain of heterogeneous bicliques}
We test our BiLPA algorithm on a type of exemplar bipartite networks, i.e. a chain network of $K$ heterogeneous bicliques connected through single nodes (Figure \ref{figure2}C). Assuming that biclique $B_i=(U_i,V_i,E_i)$ has $s_i+t_i$ nodes and $|E_i|=s_i\times t_i$ edges, then the network has a total of $N=\sum_{i=1}^K (s_i+t_i)-K+1$ nodes and $M=\sum_{i=1}^K |E_i|$ edges. The network has a clear bipartite community structure where each community corresponds to a single biclique, thus the optimal partition density is 1. Using the BiLPA algorithm described above, we can easily detect the optimal partition and identify the overlapping nodes.

\subsection{Real-world Networks}
In this subsection, we validate our algorithm on several real-world networks, including the Southern women network, the Scotland corporate interlock network, the Malaria gene-substring network, and the protein complex-drug network. We applied BiLPA to these networks with $Iter_{\max}=100$ and used $\theta$ to control the membership of each node. We can obtain a hard partition with $\theta=1.0$ and overlapping communities with $\theta<1.0$.

\textbf{The Southern women network} The famous Southern women bipartite network collected in 1930s described women's attendance to social events in the town of Natchez, Mississippi \cite{Guimer2007, Freeman2003}. It consists of 18 women and 14 events with 89 edges between them. It has become a popular benchmark for discussing and exploring bipartite networks in the social science \cite{Guimer2007}. Guimer\`{a} \cite{Guimer2007} has analyzed the modules of both women and events by three methods, including an unweighted projection method, a weighted projection method and a bipartite modularity maximization method. The unweighted projection method did not capture the true modular structure of the network. The weighted projection method and the bipartite modularity approach capture the two-community structure of the network except that one woman was partitioned wrongly. We applied BiLPA with $\theta=1.0$ to the Southern women network and identified two communities of the network, consisting of \{A1,A2,A3,A4,A5,A6,A7,A8,A9; B1,B2,B3,B4,B5,B6.B7,B8 \} and \{A10,A11,A12,A13,A14,A15,A16,A17,A18; B9,B10,B11,B12,B13,B14 \}. The corresponding partition density is 0.491.  We can also obtain four overlapping communities with $\theta=0.8$, consisting of \{A3, A4, A5, A6, A9; B3, B4, B5, B7\}, \{A1, A2, A6, A7, A8, A9; B1, B2, B3, B5, B6, B8\}, \{A10,A11,A12, A13, A16; B8, B9, B10, B12, B13, B14\}, and \{A11, A14, A15, A17, A18; B9, B10, B11\} respectively (Figure 3). The corresponding partition density is 0.601. If we use a smaller $\theta$ (e.g., $\theta=0.7$), we can also partition the network into four communities. In such cases, more nodes will be determined as overlapping ones, and the partition density will be much larger.
\begin{figure}[h]
\begin{center}
\includegraphics[width=0.8\textwidth]{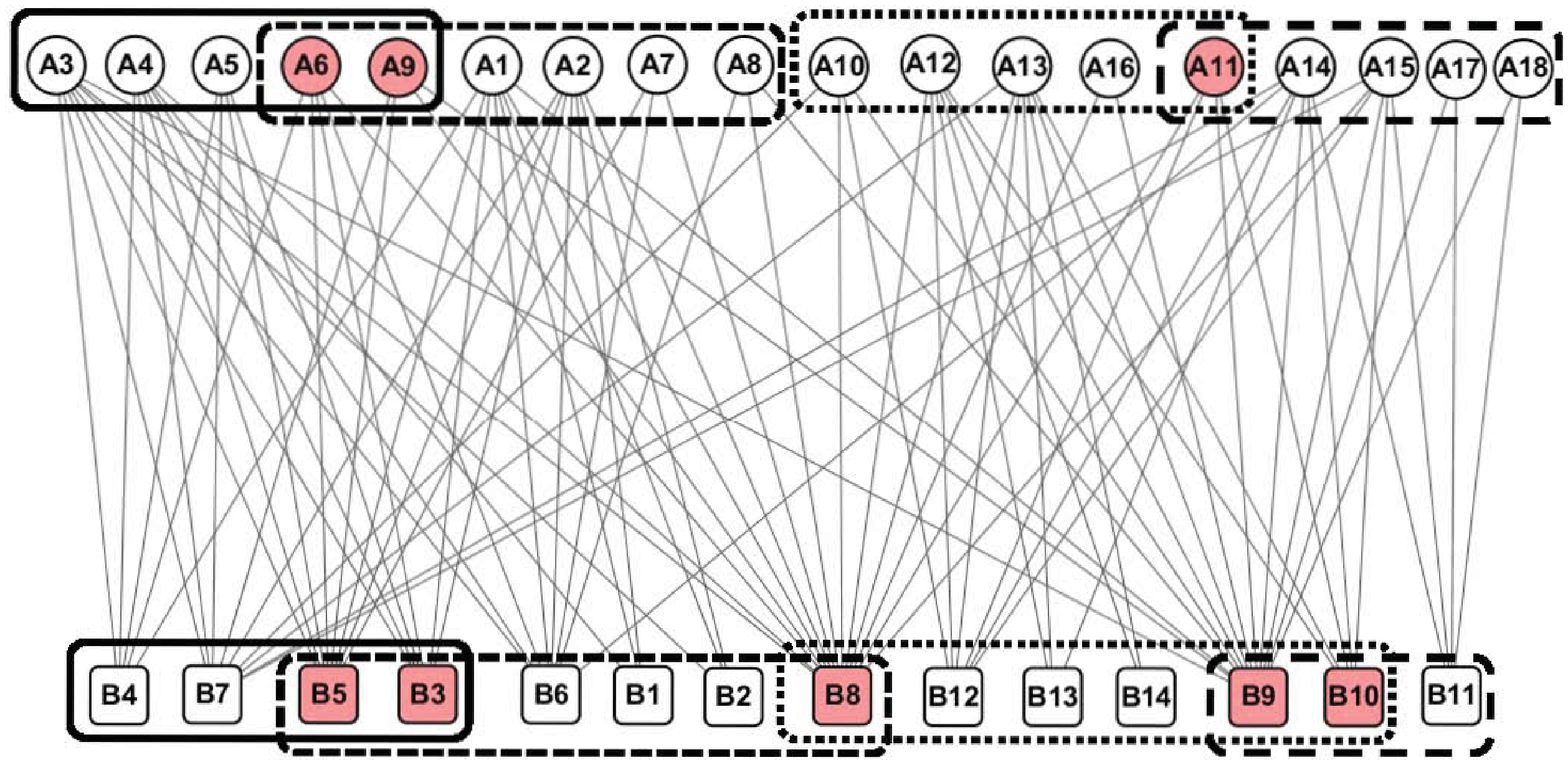}
\caption{The four overlapping communities of the Southern women network detected by BiLPA. The red nodes A6, A9, A11, B3, B5, B8, B9, B10 are overlapping ones.} \label{figure3}
\end{center}
\end{figure}%

\textbf{The Scotland corporate interlock network} The Scotland corporate interlock network consisting of 244 nodes and 356 edges describes the corporate interlocks in Scotland in the beginning of the twentieth century (1904-1905). The 244 nodes are divided into two parts, where 136 nodes represent the board members who held multiple directorships, and 108 nodes denote the firms. The edges exist between each firm and its board members. The Scotland corporate interlock network is not connected. The largest component of the network contains 131 directors and 86 firms, which form 36 communities \cite{Chen2013}.

We applied our BiLPA algorithm to the largest component of the network. We determined 42 communities (Figure 4A) and obtained a partition density 0.524 with $\theta=1.0$. We can also determine 23 nodes belong to multiple communities with $\theta = 0.8$. However, by optimizing Barber's modularity, the network can only be partitioned into 20 communities, and some small communities are merged into large communities. Obviously, optimizing our partition density with BiLPA algorithm provides higher quality community structure than other methods.

\begin{figure}[h]
\begin{center}
\includegraphics[width=0.6\textwidth]{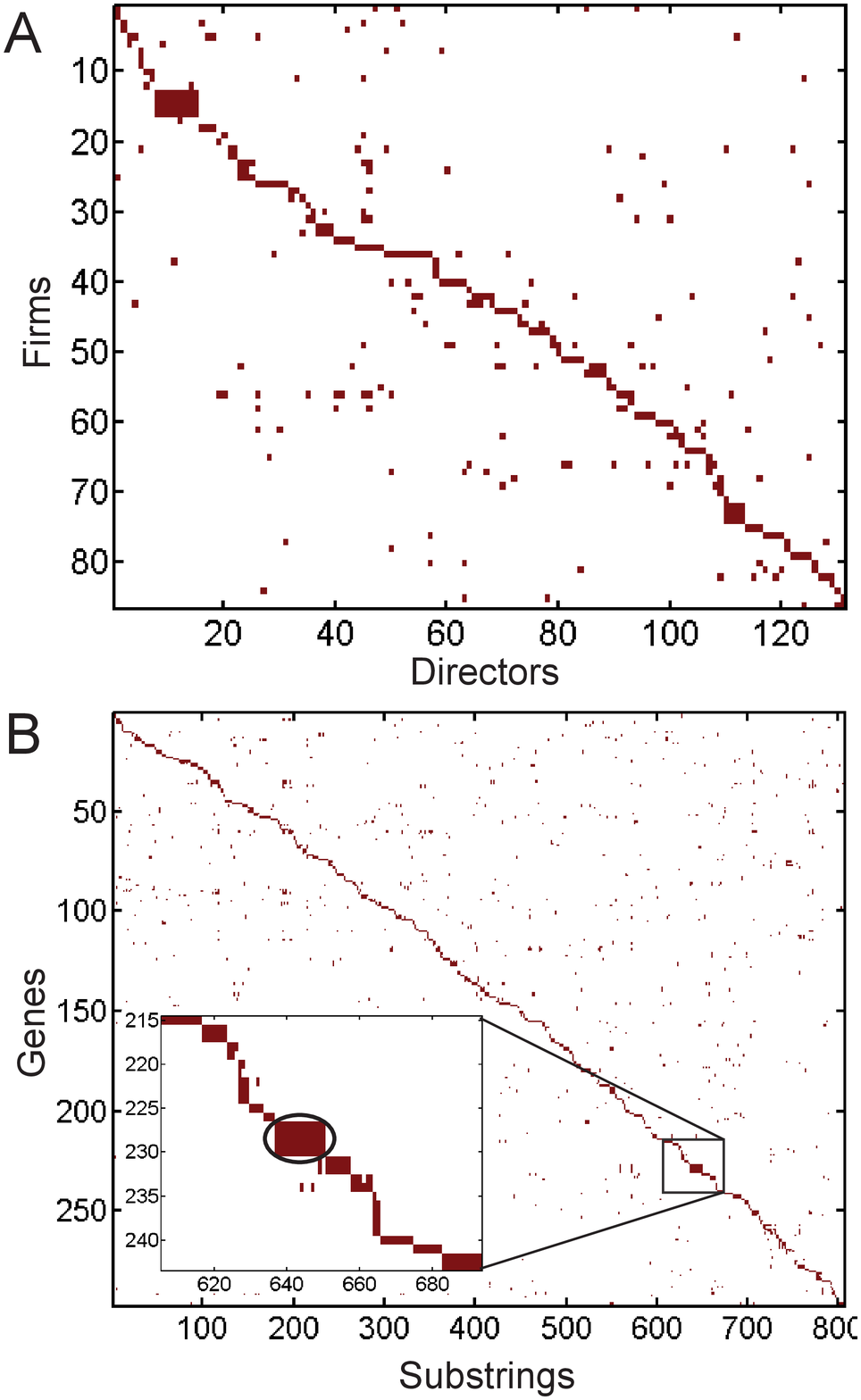}
\caption{The communities of the Scotland corporate interlock network (A) and the Malaria gene-substrings network (B) detected by BiLPA.} \label{figure4}
\end{center}
\end{figure}%

\textbf{The Malaria gene-substring network} The Malaria gene-substring network comes from the malaria parasite \emph{P. falciparum}. The parasite can evade the human immune system via a protein camouflage encoded in \emph{var} genes \cite{Larremore2013,Larremore2014}. In order to create novel camouflages, \emph{var} genes frequently recombine, which amounts to the constrained splicing and shuffling of genetic substrings, giving rise to community structures naturally \cite{Larremore2014}. The network consists of two types of nodes corresponding to 297 genes and 806 constituent substrings respectively. Each substring connects to a gene in which it is present. There are 2965 edges in the network and the link density of the bipartite network is 0.012.

In \cite{Larremore2014}, the network is partitioned into three coarse communities. We applied our BiLPA algorithm to the network and identified 158 hard communities with a maximal partition density 0.41 using $\theta=1.0$ (Figure 4B). From Figure 4B, we can see that this network is very sparse and thus partitioning it into three communities is unreasonable. Our method obtains very detailed communities. For example, BiLPA detects a biclique community consisting of four genes and 14 substring (the circled community in Figure 4B), while it was merged into a very large one in \cite{Larremore2014}. This result indicates that BiLPA can find better community structure and conquer the resolution limit showed in \cite{Larremore2014}.

\textbf{The protein complex-drug network}
The bipartite network of protein complexes and drugs investigated by Nacher and Schwartz \cite{Nacher2012} contains 1419 nodes (680 drugs and 739 complexes) and 3690 edges. Nacher and Schwartz identified network communities in both the bipartite network itself and the projected networks \cite{Nacher2012}. The number of modules in the drug projection network is 23, while the number of modules in the protein complex projection network is 17. Using the bipartite network itself, they detected 48 modules of drugs and 42 modules of complexes.

We applied our BiLPA to this network and partitioned it into 122 communities with partition density 0.57 (Figure 5A). We can see that each small disconnected subgraph is identified as one community. The giant connected subgraph is partitioned into about 90 communities. This network is very sparse and most communities are very sparsely connected with others, but densely connected within themselves. For example, we demonstrated two distinct complex-drug communities in Figure 5B and Figure 5C. The circled community in Figure 5B consists of four complexes (CORUM{\_}0441, CORUM{\_}1054, CORUM{\_}2124 and CORUM{\_}5862) and 40 drugs. Most drugs are fully connected with these four complexes. Another circled dense community in Figure 5C consists of five complexes (CORUM{\_}1223, CORUM{\_}2242, CORUM{\_}4158, CORUM{\_}5189, CORUM{\_}5526) and 16 drugs. The bipartite communities reveal that some drugs target to a same set of biological complexes. This is valuable for prediction of the detailed mechanisms of unknown ones from the well-known drugs' functions.

\begin{figure}[h]
\begin{center}
\includegraphics[width=0.80\textwidth]{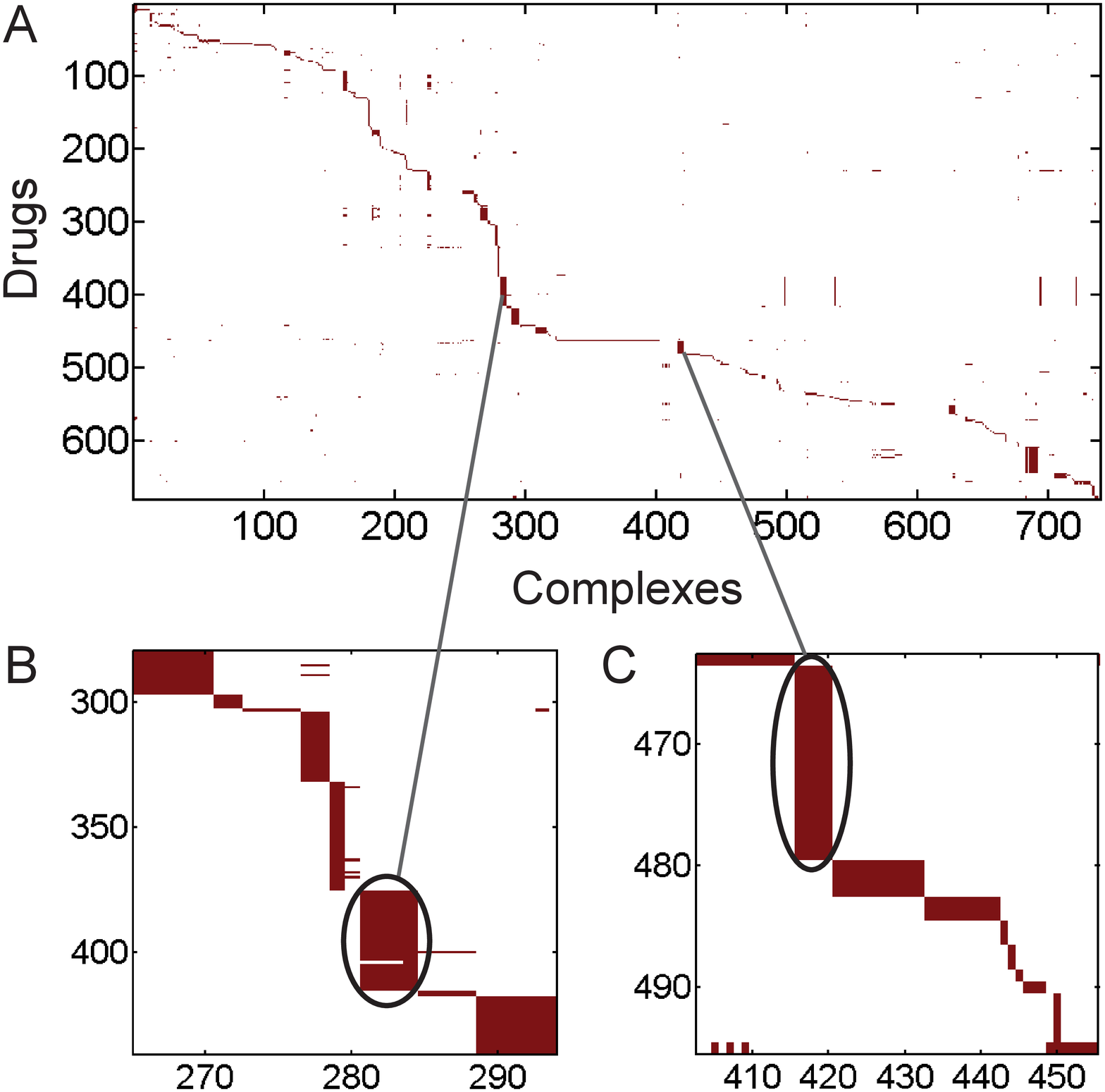}
\caption{The communities of the protein complex-drug network detected by BiLPA (A) and two illustrative community examples in (B) and (C). These heatmaps were depicted based on the rearranged matrices according to the output from BiLPA} \label{figure4}
\end{center}
\end{figure}%

\section{Conclusion and discussion}
Community structure is one of the main characteristics of bipartite networks and is valuable for understanding the organization and functions of such networks. In this paper, we propose a new quantitative function --- bipartite partition density --- for evaluating the community partition of bipartite networks. We show that our quantitative function can conquer the serious resolution limit issues of modularity-based functions. We first formulate the community detection problem for bipartite networks into an integer nonlinear programming model. Furthermore, we design a fast heuristic algorithm BiLPA for solving the community detection problem and conduct validation experiments on some artificial and real-world networks. The computational results demonstrate that our model and algorithm can detect overlapping communities in bipartite networks. Moreover, the quantitative function proposed in this paper can be easily extended to the community detection in weighted bipartite networks.

\section*{Acknowledgments}
This project was supported by the National Natural Science Foundation of China, No. 61379092, 61422309 and 11131009, the Foundation for Members of Youth Innovation Promotion Association, CAS. The Outstanding Young Scientist Program of CAS, the Scientific Research Foundation for ROCS, SEM, the Key Laboratory of Random Complex Structures and Data Science at CAS, the Funding Project for Academic Human Resources Development in Institutions of Higher Learning Under the Jurisdiction of Beijing Municipality (CIT\&TCD20130327).

\noindent \textbf{Appendix A: Community Partition Density Improves Resolution Limits}

{\bf A ring of bicliques} The ring of bicliques has a distinct community structure where each biclique corresponds to a single community, but the correct result cannot be obtained by optimizing Barber's bipartite modularity $Q$. If we partition the network into $s$ communities, and each community corresponds to a single biclique, then Barber's bipartite modularity $Q$ is
\begin{equation*}
\begin{array}{l}
\displaystyle Q_{s}=\left(\frac{mn}{s(mn+1)}-\frac{mn+1}{s(mn+1)}\frac{mn+1}{s(mn+1)} \right) \cdot s\\
\displaystyle                                                   \\
\displaystyle \quad \quad \quad =\frac{mn}{mn+1}-\frac{1}{s}.
\end{array}
\end{equation*}

If we partition the network into $\frac{s}{2}$ communities, and each community consists of two adjacent bicliques, then Barber's bipartite modularity $Q$ is
\begin{equation*}
\begin{array}{l}
\displaystyle Q_{\frac{s}{2}}=\left(\frac{2mn+1}{s(mn+1)}-\frac{2mn+2}{s(mn+1)}\frac{2mn+2}{s(mn+1)}\right)\cdot \frac{s}{2}\\
\displaystyle                \\
\displaystyle \quad \quad \quad =\frac{2mn+1}{2(mn+1)}-\frac{2}{s}.
\end{array}
\end{equation*}

When $s> 2(mn+1)$, we have
\begin{equation*}
\begin{array}{l}
\displaystyle Q_{s}-Q_{\frac{s}{2}}=\frac{mn}{mn+1}-\frac{1}{s}-\frac{2mn+1}{2(mn+1)}+\frac{2}{s}\\
\displaystyle    \\
\displaystyle  \quad \quad \quad =\frac{1}{s}-\frac{1}{2(mn+1)}\\
\displaystyle    \\
\displaystyle  \quad \quad \quad < 0.
\end{array}
\end{equation*}

The partition density $D_{s}$ of the natural partition can be easily and analytically calculated as follows:
$$D_{s}= \frac{1}{s(mn+1)}\left(\frac{(mn)^2}{mn} \cdot s\right)=\frac{mn}{mn+1}.$$
On the other hand, the partition density $D_{k}$ of the partition in which the $k$ consecutive bicliques are considered as a single community is$$D_{\frac{s}{k}}=\frac{1}{s(mn+1)} \left(\frac{(kmn+k-1)^2}{(km)(kn)} \cdot \frac{s}{k}\right)=\frac{(kmn+k-1)^2}{k^3mn(mn+1)}.$$
Then,
\begin{equation*}
\begin{array}{l}
\displaystyle D_{s}-D_{\frac{s}{k}}=\frac{mn}{mn+1}-\frac{(kmn+k-1)^2}{k^3mn(mn+1)}\\
\displaystyle   \\
\displaystyle  \quad \quad \quad =\frac{k^3(mn)^2-(kmn+k-1)^2}{k^3mn(mn+1)}\\
\displaystyle  \\
\displaystyle  \quad \quad \quad=\frac{(mn)(k-(1+\frac{k-1}{kmn})^2)}{k(mn+1)}.
\end{array}
\end{equation*}
Supposing $mn\geq 2$, $s\geq 4$, when $k=2$,
\begin{equation*}
\begin{array}{l}
\displaystyle D_{s}-D_{\frac{s}{2}}=\frac{(mn)(2-(1+\frac{1}{2mn})^2)}{2(mn+1)}\\
\displaystyle \quad \quad \quad>0.
\end{array}
\end{equation*}
When $k\geq 3$,
\begin{equation*}
\begin{array}{l}
\displaystyle D_{s}-D_{\frac{s}{3}}=\frac{(mn)(k-(1+\frac{k-1}{kmn})^2)}{k(mn+1)}\\
\displaystyle \\
\displaystyle \quad \quad \quad>\frac{(mn)(k-(1+\frac{1}{mn})^2)}{k(mn+1)}\\
\displaystyle \\
\displaystyle \quad \quad \quad \geq \frac{(mn)(k-(1+\frac{1}{2})^2)}{k(mn+1)}\\
\displaystyle \\
\displaystyle \quad \quad \quad>0.
\end{array}
\end{equation*}

{\bf A bipartite network consisting of bicliques of different sizes} Suppose there is a bipartite network consisting of four bicliques, two of which are $B(n,n)$, and the other two are $B(m,m)$, for $2\leq m\leq n $. This bipartite network has a natural community structure of four communities, where each biclique corresponds to one community. If we partition the bipartite network by optimizing Barber's bipartite modularity $Q$, we will obtain three communities, where two small bicliques merge into one community. We can calculate their corresponding $Q$ as follows:
\begin{equation*}
\begin{array}{l}
\displaystyle Q_{separate}=(\frac{n^2}{2n^2+2m^2+3}-\frac{n^2+1}{2n^2+2m^2+3}\frac{n^2}{2n^2+2m^2+3}) \\
\displaystyle \quad  +(\frac{n^2}{2n^2+2m^2+3}-\frac{n^2+1}{2n^2+2m^2+3}\frac{n^2+1}{2n^2+2m^2+3})\\
\displaystyle \quad +(\frac{m^2}{2n^2+2m^2+3}-\frac{m^2+1}{2n^2+2m^2+3}\frac{m^2+1}{2n^2+2m^2+3})\\
\displaystyle \quad +(\frac{m^2}{2n^2+2m^2+3}-\frac{m^2+1}{2n^2+2m^2+3}\frac{m^2}{2n^2+2m^2+3}).\\
\end{array}
\end{equation*}

\begin{equation*}
\begin{array}{l}
\displaystyle Q_{merge}=(\frac{n^2}{2n^2+2m^2+3}-\frac{n^2+1}{2n^2+2m^2+3}\frac{n^2}{2n^2+2m^2+3}) \\
\displaystyle \quad \quad \quad +(\frac{n^2}{2n^2+2m^2+3}-\frac{n^2+1}{2n^2+2m^2+3}\frac{n^2+1}{2n^2+2m^2+3})\\
\displaystyle \quad \quad \quad +(\frac{2m^2+1}{2n^2+2m^2+3}-\frac{2m^2+1}{2n^2+2m^2+3}\frac{2m^2+2}{2n^2+2m^2+3}).  \\
\end{array}
\end{equation*}

\begin{equation*}
\begin{array}{l}
\displaystyle Q_{separate}-Q_{merge}=(\frac{m^2}{2n^2+2m^2+3}-\frac{m^2+1}{2n^2+2m^2+3}\frac{m^2+1}{2n^2+2m^2+3})\\
\displaystyle \quad \quad \quad +(\frac{m^2}{2n^2+2m^2+3}-\frac{m^2}{2n^2+2m^2+3}\frac{m^2+1}{2n^2+2m^2+3})\\
\displaystyle \quad \quad \quad -(\frac{2m^2+1}{2n^2+2m^2+3}-\frac{2m^2+1}{2n^2+2m^2+3}\frac{2m^2+2}{2n^2+2m^2+3})\\
\displaystyle \quad \quad \quad =\frac{-2n^2+2m^4+m^2-2}{(2n^2+2m^2+3)^2}.
\\
\end{array}
\end{equation*}
If $n\geq m^2+1$, then $Q_{separate}-Q_{merge}<0$. This means that modularity $Q$-based optimization will merge two smaller bicliques $B(m,n)$s into one community.

In the following, we prove that partition density $D$-based optimization does not have this resolution limit.
\begin{equation*}
\begin{array}{l}
\displaystyle D_{separate}=\frac{1}{2n^2+2m^2+3}(n^2+n^2+m^2+m^2) \\
\displaystyle \quad \quad =\frac{2n^2+2m^2}{2n^2+2m^2+3}.\\
\end{array}
\end{equation*}

\begin{equation*}
\begin{array}{l}
\displaystyle D_{merge}=\frac{1}{2n^2+2m^2+3}(n^2+n^2+\frac{(2m^2+1)^2}{4m^2}).\\
\end{array}
\end{equation*}

It is easy to verify that when $m\geq 2$,
\begin{equation*}
\begin{array}{l}
\displaystyle D_{separate}-D_{merge}=\frac{(2m^2-1)^2-2}{4m^2(2n^2+2m^2+3)}>0.
\end{array}
\end{equation*}

\end{document}